# The Impact of AI on Educational Assessment: A Framework for Constructive Alignment

Patrick Stokkink
Faculty of Technology, Policy and Management, TU Delft, Netherlands Contact: p.s.a.stokkink@tudelft.nl

**Abstract**

The influence of Artificial Intelligence (AI), and specifically Large Language Models (LLM), on education is continuously increasing. These models are frequently used by students, giving rise to the question whether current forms of assessment are still a valid way to evaluate student performance and comprehension. The theoretical framework developed in this paper is grounded in Constructive Alignment (CA) theory and Bloom's taxonomy for defining learning objectives. We argue that AI influences learning objectives of different Bloom levels in a different way, and assessment has to be adopted accordingly. Furthermore, in line with Bloom's vision, formative and summative assessment should be aligned on whether the use of AI is permitted or not.

Although lecturers tend to agree that education and assessment need to be adapted to the presence of AI, a strong bias exists on the extent to which lecturers want to allow for AI in assessment. This bias is caused by a lecturer's familiarity with AI and specifically whether they use it themselves. To avoid this bias, we propose structured guidelines on a university or faculty level, to foster alignment among the staff. Besides that, we argue that teaching staff should be trained on the capabilities and limitations of AI tools. In this way, they are better able to adapt their assessment methods.

# 1 Introduction

Over the last years, the use of Artificial Intelligence (AI) has peaked. The most widely known models are Large Language Models (LLMs), an AI system trained on large amounts of textual data that enable it to perform a wide range of language-related tasks like translation, summarization, and conversation, and Generative AI (Gen-AI) models that are designed to create new content—such as text or images by learning patterns from large datasets. Especially after the introduction of ChatGPT by OpenAI (OpenAI, 2022) in November 2022, the interest in AI has peaked, as depicted in Figure 1. Already before that, AI has started to be integrated in many aspects of society such as healthcare (Shaheen, 2021), criminal justice (Završnik, 2020) and marketing (Verma et al., 2021; Mariani et al., 2022).

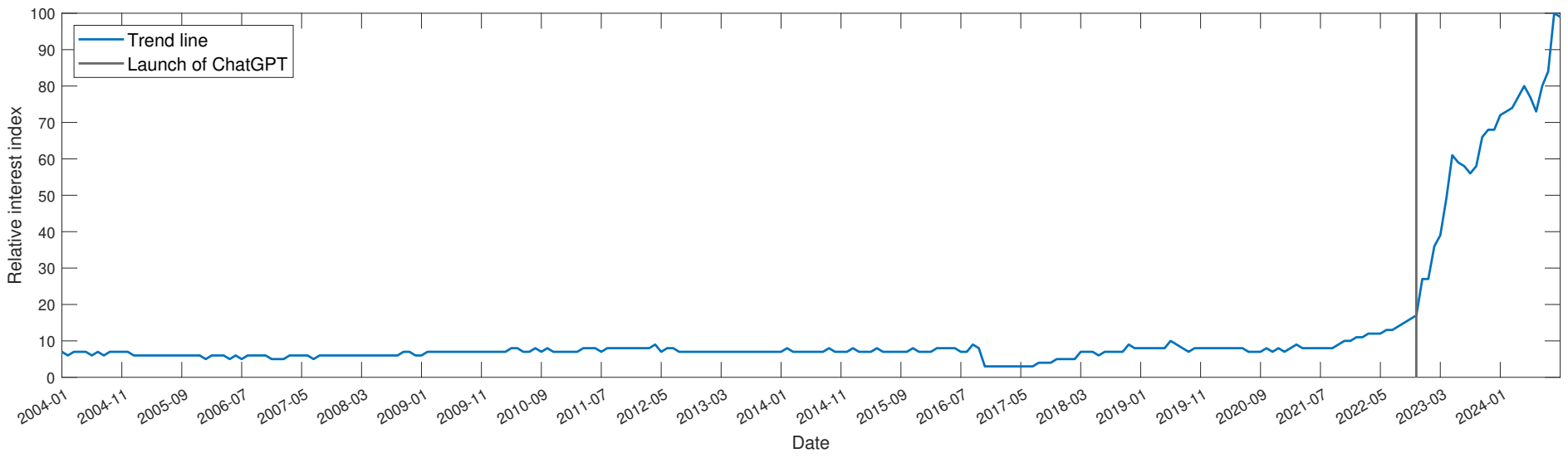

Figure 1: Relative interest over time for "Artificial Intelligence" according to Google Trends (2024)

Naturally, AI has also started influencing academia and education. In academia, AI can aid researchers in performing literature reviews (Tomczyk et al., 2024), writing (Imran and Almusharraf, 2023; Fang et al., 2023), coding (Poldrack et al., 2023; Huang et al., 2019), and many more. As a consequence, a concern arises whether this is in conflict with academic integrity (Bin-Nashwan et al., 2023). In education, AI has influenced the sector from both sides. Fahimirad et al. (2018); Ng et al. (2023) provide reviews on the use of AI in teaching and learning. On the one hand, teachers can use AI, for example in the development of teaching material (Williyan et al., 2024; Koraishi, 2023), providing feedback (Nysom, 2023), and student grading (Kumar, 2023; Jonäll, 2024). Goel and Joyner (2017) even went as far as using AI in the development of an online course on AI. On the other hand, students use AI to help them study (Lai, 2021; Wu et al., 2022; Wang et al., 2024) and to complete assignments (Bolboacă, 2023; Fyfe, 2023).

In this paper, we focus on a problem that has been initiated by the latter concept. The use of AI by students has made it more and more complicated for teachers to assess their students. In this paper we answer the following research questions:

- Does AI influence whether a certain student assessment is valid?
- How is the influence of AI on the assessment of Learning Objectives (LOs) associated to different Bloom levels?

We highlight that the concept of assessment in this paper is different than that in González-Calatayud et al. (2021), where AI is used as a tool to help teachers in grading student assignments and providing feedback to students.

The remainder of this paper is organized as follows. The theoretical framework and corresponding findings, based on Constructive Alignment and Bloom's taxonomy, are discussed in Section 2. A survey has been conducted among course coordinators on the influence of AI on student assessment, which is presented in Section 3. In Section 4, the paper is concluded and a set of concrete recommendations are given based on the findings in this paper.

## 2 Theoretical framework

The foundation of this paper lies in Constructive Alignment (CA) as introduced by Biggs (1996). This educational design framework is highly student centred and aligns three critical elements of learning: learning objectives (what students should know and be able to do), teaching and learning activities (what students will engage in to achieve these objectives), and assessment methods (how students' attainment of objectives will be measured). This concept is typically depicted by a triangle, as shown in Figure 2. In this paper, the focus will be on the alignment of learning objectives with assessment methods.

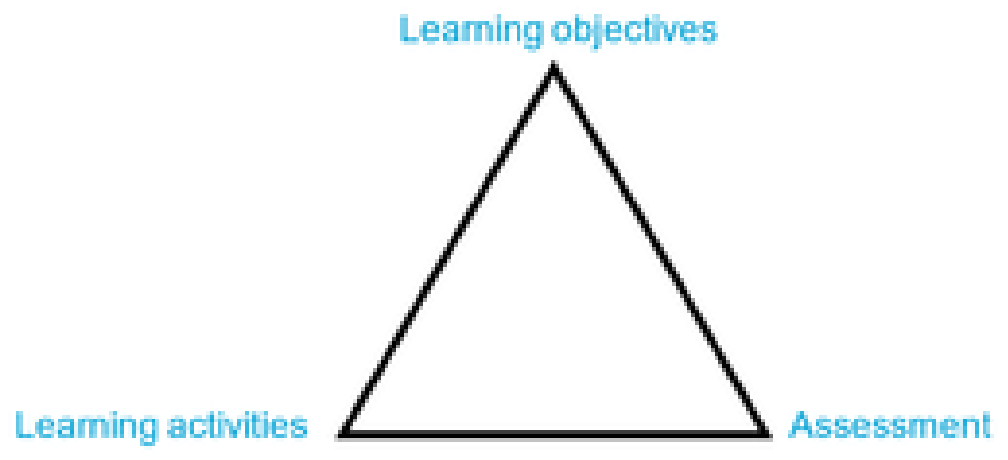

Figure 2: Constructive alignment triangle (TU Delft, 2024)

For the definition of the learning objectives, we consider those developed according to the standard set by Bloom and Krathwohl (1956) and revised by Krathwohl (2002), referred to as Bloom's taxonomy. According to Bloom's taxonomy, learning objectives can be classified into six sets of objectives of increasing level of complexity. The revised taxonomy including a one-line explanation of the class as proposed by Krathwohl (2002), is as follows:

1. Remember: Retrieving relevant knowledge from long-term memory.
2. Understand: Determining the meaning of instructional messages, including oral, written, and graphic communication.
3. Apply: Carrying out or using a procedure in a given situation.
4. Analyze: Breaking material into its constituent parts and detecting how the parts relate to one another and to an overall structure or purpose.

5. Evaluate: Making judgments based on criteria and standards.
6. Create: Putting elements together to form a novel, coherent whole or make an original product.

We first discuss the relation between formative and summative assessment. Then, we discuss the effect of AI on the assessment of learning objectives corresponding to each Bloom level. Finally, we discuss the different levels of AI usage to be considered. This theory is then evaluated for a diverse set of courses, learning objectives, and assessment methods.

## 2.1 Alignment of formative and summative assessment

According to Bloom et al. (1971) formative assessment supports a learning process that is more personalized, adaptive, and effective in helping students reach their full potential. Formative assessment takes place through the course to help students learn and practice and is not graded. Summative assessment, on the other hand, is used to assess a student's performance and is graded. Alignment of formative and summative assessment is crucial according to Bloom et al. (1971). He argues that formative assessments serve as checkpoints where students receive feedback, identify areas for improvement, and gain additional support. Without proper alignment, these checkpoints are less useful.

In line with Bloom's vision on the alignment of formative and summative assessment, we propose the following:

**Proposition 1.** *AI is forbidden/discouraged/allowed/encouraged in summative assessment* if and only if *it is forbidden/discouraged/allowed/encouraged in formative assessment.*

This proposition follows the same intuition as the use of other supporting material with which teachers have dealt for decades or centuries, such as the use of dictionaries and calculators. With respect to assessing mathematics, allowing the use of a calculator during a formative assessment does not prepare the student well for a summative assessment where the use of a calculator is forbidden. Similarly, with respect to assessing language comprehension, allowing the use of a dictionary during a formative assessment does not prepare the student well for a summative assessment where the use of a dictionary is forbidden. In line with these classical examples, the use of AI tools should only be allowed during formative assessment if it is also allowed during summative assessment.

As an example, consider a programming course where students are asked to code a certain problem. In case students perform their formative assessment using AI tools that provide part of the code for them, the formative assessment does not provide accurate feedback and areas for improvement for a summative assessment where the use of AI is not allowed. On the other hand, if the use of AI tools is allowed or even encouraged during the summative assessment, it would be useful for the students to also practice with the use of these tools during their formative assessment. In this way, they can obtain feedback and find areas for improvement.

A remark has to be made for this statement, regarding the different purposes for which AI can be used. According to, among others, Lai (2021); Wu et al. (2022); Wang et al. (2024) AI can form a useful aid for students in studying. This can form a counter argument on the use of AI in formative assessment, even though it is not allowed in summative assessment. Even more so, the use of AI tools in Learning Activities (LAs) may be considered useful even though it is not allowed in summative assessment. The complexity of this counter argument lies in the confusion that may arise among students concerning whether the use of AI is accepted or not. Because of this, this argument has to be treated with care.

## 2.2 Influence of AI on the assessment of different Bloom levels

We distinguish between the six Bloom levels as discussed in the revised taxonomy by Krathwohl (2002). Here, we emphasize that in this paper we focus on the influence of AI on the validity of assessment. Whether LOs are still relevant in the presence of AI is another extremely relevant research direction that falls outside the scope of this work. Views of this can be found, for example, in Oregon State University (2024), where Bloom's taxonomy is revisited based on how generative AI can supplement learning.

1. **Remember**: Learning objectives in this level focus on retrieving relevant knowledge from long-term memory. To assess this, a written or oral exam can be used, for example multiple-choice, true-false, or listing exercises. Given that students are assessed on whether they can retrieve information from their memory, the use of AI should be *forbidden* to properly assess the attainment of this learning objective. Typically, this can be easily enforced through examinations with limited access to software and internet or traditional written examinations.

2. **Understand**: Learning objectives in this level focus on having an understanding of instructional messages. Questions to test for understanding typically start with "In your own words, ...", which already signals that the use of AI tools to answer such questions is disruptive for assessing attainment of the learning objective. Questions typically involve paraphrasing, summarizing or giving examples, which are all exercises that LLMs are highly capable of. Clearly, the use of AI in summative assessment of these learning objectives should therefore be *forbidden* to properly assess the attainment of this learning objective. Similar to the Remember-level, this can be easily enforced through examinations with limited access to software and internet or traditional written examinations.

3. **Apply**: Learning objectives in this level focus on carrying out a procedure in a given situation. The influence of AI on assessment of this level is not as straightforward as for the previous two levels and may depend on the field of studies. When a student is asked to apply a simple formula, this can be done through a written examination where attainment of the level can be properly assessed. However, when students are asked to apply more complex methods or frameworks, this is typically assessed through an assignment where students can work outside the classroom. In this case, monitoring the use of AI is more complicated. Also, whether the use of AI should be *forbidden, discouraged, allowed or encouraged* in this case is not obvious and typically requires the *judgement of the course coordinator*. Examples of this can be found in the case studies. In case AI is forbidden during a take-home assignment, a possible approach to assess the attainment of the learning objective is to have an oral or written examination where students explain their procedure and reasoning while working on the take-home assignment.

4. **Analyze, Evaluate, Create**: For the final three levels, the intuition is highly similar to that of the apply level. For analyzing, evaluating or creating smaller problems, solution approaches or designs, a written examination can be used to guarantee that AI is not used by the students while completing the assessment. For larger problems, systems or designs, assignments are typically used as an assessment method. In this case, combining the assignment with an oral or written examination where students explain their procedure and reasoning can again form a useful addition. Similarly, it is up to the course coordinator to what extent AI can be used during the assessment. In this, one may decide that AI can be used as an assistant. However, it should be noted that if students are assessed on their capabilities for critical thinking or innovative designs, this should be executed by the students themselves, rather than by AI tools.

## 2.3 Different levels of AI usage

Although the recommendations on the use of AI in assessment for the first two Bloom levels are rather concrete, the judgement of the course coordinator is needed for the latter four levels. To provide a decision-support tool for course coordinators, we consider four levels of AI usage, and we consider that an assessment is typically made up of various components. Here, we focus on assessments where students are not continuously supervised, such as take-home assignments. For assessments where students are continuously supervised, such as written examinations, the previous subsection provides concrete recommendations.

We consider that an assessment $A$ is typically made up out of multiple components $C$ and that the desired level of AI usage may differ for each of these components. We classify the use of AI into the following four levels, where the ordering is based on an increasing reliance on AI:

0. The student completes the entire assessment without the use of AI.
1. The student completes component C him/herself and uses AI to provide feedback.
2. The student uses AI to complete component C and verifies him/herself whether the output is correct.

3. The student uses AI to complete component C and does not verify him/herself whether the output is correct.

Whereas for one component the use of AI may be permitted, it may not be permitted for another component. Consider the following arbitrary example: $A$ = Write a report describing your analysis of a system $X$, $C$ = Convert the analysis into a report. In this case, the defined AI levels may look like this:

0. The student writes their own report, no AI is used to enhance the use of English afterwards.
1. The student writes their own report, AI is used to enhance the use of English afterwards.
2. AI is used to generate a paragraph of text for a report, the student evaluates him/herself whether the output is correct.
3. AI is used to generate a paragraph of text for a report, the student directly copies it into the report.

Assume the learning objective is: "At the end of this course, the student should be able to analyze a system $X$." In this case, the course coordinator may decide to accept all AI levels 0, 1, and 2. Now assume the learning objective is: "At the end of this course, the student should be able to write a report according to academic standards". In this case, AI level 0 is acceptable, whereas AI levels 1 and 2 are debatable. In either case, AI level 3 has been marked as unacceptable, given that students are always responsible for their own work and should therefore always verify the correctness of the AI-generated material they include in their report.

# 3 Survey

A survey was conducted under the course coordinators of the Technical University of Delft to evaluate their awareness of the influence of AI on assessment. In addition to this, the view of lecturers on AI in general and the development of university policies was investigated. The survey was open to all faculties, but was advertised specifically for staff members at two faculties to get a diverse set of perspectives. A total of 24 complete responses were collected. Out of these, 13 responses were from the TPM faculty, 8 responses were from the BK faculty and the remaining responses were from other faculties. In terms of level of education, 9 responses were filled for a bachelor's course, 14 for a master's course, and 1 for a doctoral course. The majority of the responses considered a take-home assignment (77%) and the others considered an in-class written assignment.

## 3.1 Perception of lecturers

We first evaluate the general perception of lecturers towards AI and the use thereof. Lecturers were presented a set of statements for which they could indicate to what extent they agree. The results are displayed in a box plot in Figure 3. Teachers are unanimous on the fact that education and assessment should be adapted to AI and to the fact that students should be taught how to use AI properly. Interestingly, most lecturers recognize they could benefit from using AI in their work, but a significantly smaller share of them actually use AI on a daily basis. Many, but not all, lecturers encourage students to use AI to correct their English and a smaller share encourages students to use AI for brainstorming whereas discouraging the use of AI in graded assignments is generally not favored.

An interesting correlation can be observed between AI usage by teachers and to what extent they want students to use AI. The responses to the first statement have a significant positive correlation with statement 5 and 6 and a significant negative correlation with statement 3. This also marks an important risk of allowing lecturers to determine their own AI regulations. Our results indicate that lecturers are generally biased based on their own usage of AI.

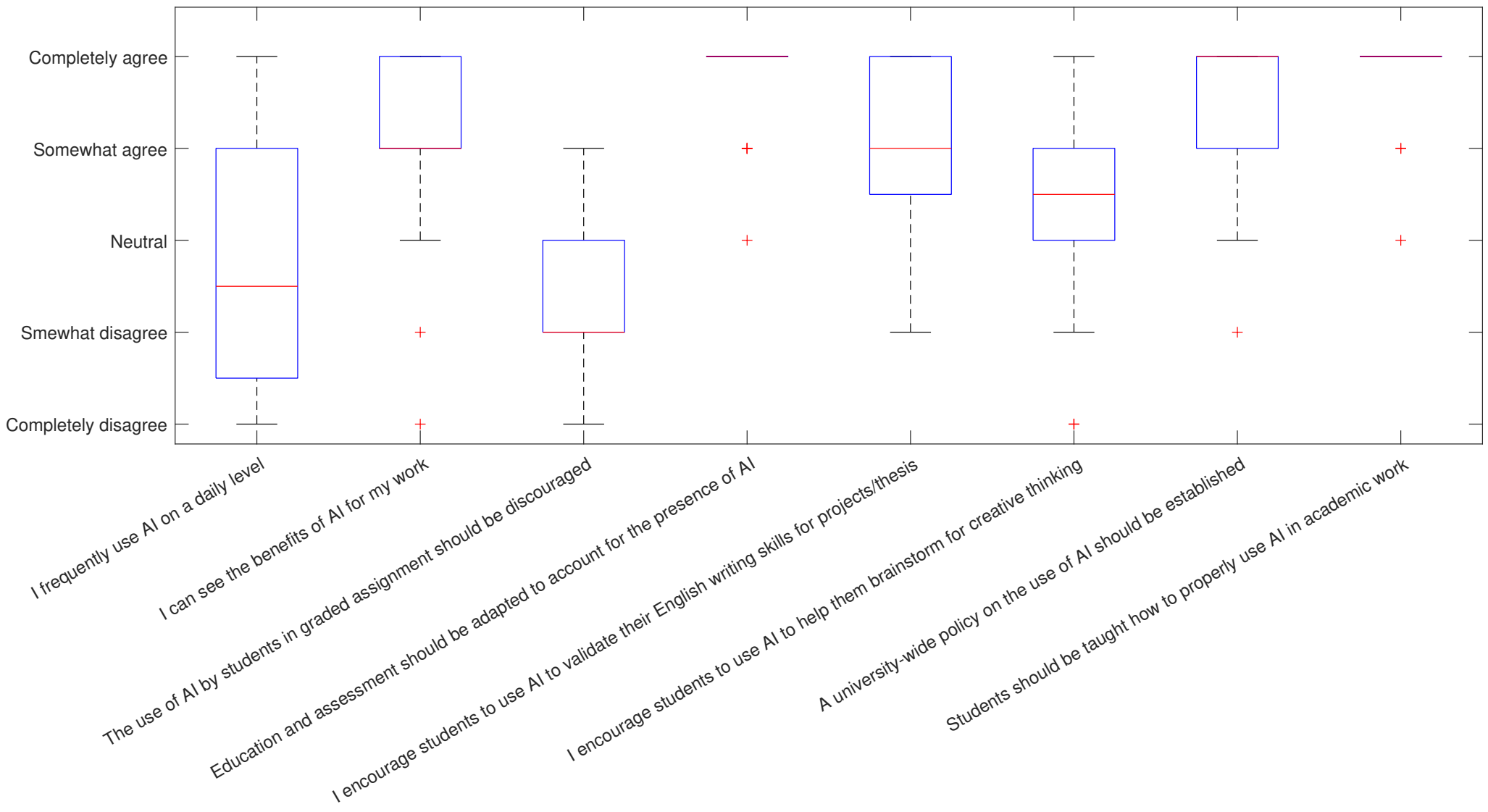

Figure 3: Perception of lecturers

## 3.2 Bloom level, AI level and implementation difficulty

We asked the lecturers to comment on their learning objectives, assignments, and the desired AI levels. Lecturers were encouraged to decompose every assignment into components as described in Section 2.3, but some of them only reported a single AI level for the entire assignment. Lecturers were also asked to report the Bloom level of their learning objectives. Nevertheless, not all lecturers managed to report the Bloom level. In that case, we filled in the Bloom level manually in case it was easy to deduce. If not, we removed the response from consideration, leaving us with 16 useful observations.

The responses of teachers have been classified by Bloom level. As courses and assignments can have multiple learning objectives with varying Bloom levels, we consider the minimum, maximum and average level. The desired AI level by the lecturer and the difficulty of enforcing this level are displayed in Figure 4. The results indicate that for higher Bloom levels, teachers generally allow for higher AI levels. For lower Bloom levels, AI usage is generally not desired. This is in line with the theoretical findings in Section 2.2.

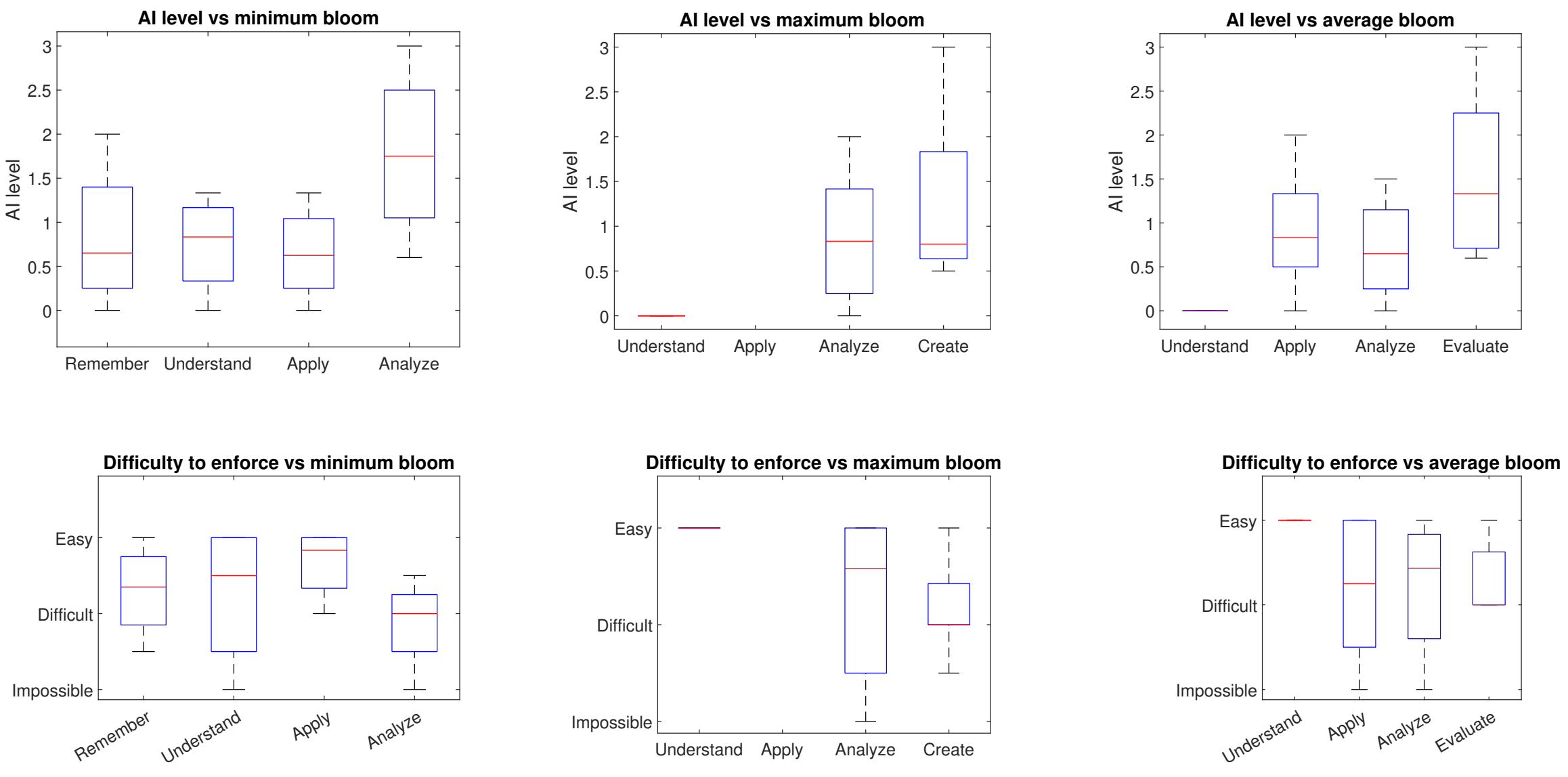

Figure 4: AI level and difficulty to enforce for varying Bloom levels

The difficulty to enforce a certain AI level is typically rather spread. The reason for this is embedded in the details of the specific assignment. It is noteworthy that lecturers gave highly contradicting explanations on whether or not AI was jeopardizing the quality of their assessment. Whereas some lecturers indicated that they believe they could easily recognize the use of AI when reading a report, others indicate it is barely possible to identify this. The latter is also supported by research showing that people cannot properly differentiate between human-written or AI-generated poetry (Köbis and Mossink, 2021). Furthermore, some lecturers believe that when questions become more "specific", "contextualized" or consider a student's "reflection". Other lecturers recognize that also these types of questions can be answered by AI with relative ease. Some lecturers suggest the use of presentations and question and answer (Q&A) sessions to validate whether the written work of students was truly their own.

Additionally, the survey results indicate that lecturers who use AI themselves have different desired AI levels compared to those lecturers who do not use AI. Similarly, their perception of difficulty to enforce is different. This supports the previous findings about the bias of teachers. Nevertheless, to quantify these results the sample size of both groups would need to be extended.

## 4 Conclusion and discussion

With the rapid development of AI, we conclude that proper adjustment of assessment methods to align with learning objectives is essential. First of all, formative and summative assessment should be aligned and students should be informed properly about whether or not the use of AI is permitted. Furthermore, depending on the Bloom level of a learning objective and the desired level of AI usage indicated by the course coordinator, it may be appropriate to switch from take-home assignments to assessing students in a controlled environment.

Survey results indicate that the perception of lecturers highly varies and that their perception of whether the use of AI should be encouraged or not is often biased by whether they use AI themselves. Nevertheless, all of the surveyed lecturers agreed on the notion that education and assessment should be adapted to account for the presence of AI and students should be properly educated about the use of AI in academic work. Lecturers indicate that the desired AI level is different depending on the Bloom level associated with the learning objectives. This is in line with the theoretical findings in this paper.

The findings presented in this paper should be interpreted with caution, as they are based on lecturers' self-assessments of the risks associated with AI in their own evaluation methods. Since this perspective may be subject to bias or limited by individual experiences, as was clearly illustrated by the results of the survey, further validation of these results is necessary. Future research should incorporate input from students and external experts to provide a more comprehensive and balanced understanding of AI's impact on assessment.

Based on the analysis presented in this paper, we propose the following concrete recommendations for educational staff:

- Firstly, the alignment between assessment methods and learning objectives has become even more critical with the increasing prevalence of AI in education. As AI tools can assist students in various ways, it is essential to design assessments that accurately measure the intended learning outcomes while considering the potential influence of AI-generated content.
- Secondly, clear communication regarding the permitted use of AI should be established. Students must be explicitly informed about whether, and to what extent, AI-generated content is allowed in assignments, examinations, and other coursework. This transparency will help prevent misunderstandings and ensure that students adhere to the intended academic integrity standards.
- Thirdly, while the implementation of a university-wide policy on AI may be challenging, given the significant variations across disciplines, courses, and even individual learning objectives, it remains crucial to establish clear and structured guidelines.

- Finally, teaching staff should receive adequate training and information regarding the capabilities and limitations of AI tools. A well-informed faculty will be better equipped to make informed decisions about integrating AI into their teaching and assessment practices. Moreover, ensuring alignment among educators within a program or institution will help maintain consistency in expectations and assessment criteria, ultimately contributing to a fair and effective learning environment.